\documentclass[12pt]{article} 

\newcommand{\COMMENTO}[1]{}

\newcommand{\spazio}{{\,\,\,\,\,}}

\newcommand{\calE}{{\cal E}}
\newcommand{\calF}{{\cal F}}
\newcommand{\calG}{{\cal G}}
\newcommand{\calP}{{\cal P}}

\newcommand{\dslash}{{\slash \kern -1ex \partial}}

\def\C{{\rm C}}

\def\Z{{\rm Z}}

\def\Ft{{F^t}}

\begin{document}

\begin{titlepage}
\rightline{DFTT 10/05}
\rightline{\hfill April 2005}
\vskip 1.2cm
\centerline{\Large \bf
Boundary states for branes with non trivial homology 
}
\centerline{\Large \bf
 in constant closed and open background.
}

\vskip 1.2cm

\centerline{\bf Igor Pesando\footnote{e-mail:
ipesando@to.infn.it}}
\centerline{\sl Dipartimento di Fisica Teorica, Universit\`a di
Torino}
\centerline{\sl Via P.Giuria 1, I-10125 Torino, Italy}
\centerline{\sl and I.N.F.N., Sezione di Torino}

\begin{abstract}
For the bosonic string on the torus
we compute boundary states describing branes with not trivial homology
class in presence of constant closed and open background.
It turns out  that boundary states with non trivial open
background generically require the introduction of non physical
``twisted'' closed sectors, that only $F$ and not ${\cal F}=F+B$
determines the geometric embedding for $Dp$ branes with $p<25$ and
that closed and open strings live on different tori which are
relatively twisted and shrunk.
Finally we discuss the T-duality transformation for the open string in
a non trivial background.

\end{abstract}

\end{titlepage}

%

\section{Introduction.}
Systems of interacting open- and closed-strings
play important roles in several aspects of string theory.
It has been found out that
theories even formulated as pure closed-strings
have open-string sectors which can be described as boundary states.
One of the most important features of
such open-closed mixed systems is a duality
between open- and closed-strings.
This duality becomes manifest, for instance,
by seeing one-loop diagrams of open-string
as tree propagations of closed-string
through modular transformations
on the string world-sheets.
In the systems of D-branes,
this duality should be a rationale for
correspondence between gauge theory on the world-volumes
(open-string sector) and gravity theory in the bulk space-time
(closed-string sector).
AdS/CFT correspondence
may be regarded as one of the most remarkable examples
of such bulk-boundary correspondence (see for example
\cite{DiVecchia:2005vm} for a review along these lines).

It is well-known that the so called boundary states
provide a description of D-branes in closed string theory.
Open  and closed strings interact on the branes
in space-time and therefore boundary states can be used to
describe these interactions.
In \cite{Murakami:2002yd} this was achieved for tachyons and gluons
in a non compact spacetime with a constant $B$ field background while
in \cite{Pesando:2003ww} for all open states in a compact spacetime
without background fields.

On the other hand intersecting branes on tori have been used as a
building block for constructing (semi)realistic versions of the standard model
(see for example \cite{Uranga:2003pz,Kiritsis:2003mc}
 for  reviews).
This construction requires considering branes with non trivial homology in presence
of both open and closed background fields  since the geometric
embedding of these branes is governed by these ``parameters'',
i.e. background fields values.

In this article we want to build boundary states for bosonic boundary
states describing branes with non trivial homology on tori in presence
of constant open and closed background fields (see also
\cite{Arfaei:1999jt,Kamani:2002bb,Kamani:2002ib,Kamani:2001cg} for related 
construction).
This paper is organized as follows.
In section 2 we fix our conventions and review the closed string 
quantization on a torus in presence of background fields and  the
action of T-duality. In this section we describe also the quantization
of open string  describing homologically non trivial branes in this
closed background in presence of a constant magnetic field.
In section 3 we proceed to construct the boundary states for the $D25$
brane: we find that it is necessary to introduce non physical
``twisted'' states to accomplish this task.
In section 4 we discuss the action of T-duality on these states from
the closed string point of view and we sketch how this can be derived
from the open string point of view ( see \cite{Pesando:2005mah} for
more details)
Finally in section 5 we draw our conclusions.

\section{Review of quantization of string and T-duality.}
\subsection{The quantization of closed string in a constant $G$ and
  $B$ background.}
In order to fix our notation and conventions we start writing the
usual action for the closed string
\begin{eqnarray*}
S
&=&
-\frac{1}{4\pi\alpha'}\int d\tau\, \int^\pi_0 d\sigma 
\left(
\sqrt{-g} g^{\alpha\beta} G_{\mu\nu} 
\partial_\alpha X^\mu \partial_\beta X^\nu
-\epsilon^{\alpha\beta} B_{\mu\nu}
\partial_\alpha X^\mu \partial_\beta X^\nu
\right)
\nonumber\\
&=&
\frac{1}{4\pi\alpha'}\int d\tau\, \int^\pi_0 d\sigma 
\left(
G_{\mu\nu} \left( \dot X^\mu \dot X^\nu - X'^\mu X'^\nu \right)
-2 B_{\mu\nu} \dot X^\mu  X'^\nu
\right)
\end{eqnarray*}
where we have chosen the conformal gauge $g_{\alpha \beta}\propto
\eta_{\alpha \beta}=diag(-1,1)$ and chosen $\epsilon^{0 1}=-1$.
We suppose moreover that the dimensionful 
background fields $G_{st}=||G_{\mu\nu}||$ 
and $B_{st}=||B_{\mu\nu}||$ be constant, 
$G_{i 0}=B_{i 0}=0$ ($i=1,\dots d$) and
the spacial  coordinates
$X=|| X^i ||$  be periodic of period $2\pi$
\begin{eqnarray}
 X^i &\equiv& X^i + 2\pi 
\label{X_closed_periodicity}
\end{eqnarray}
and  dimensionless.
From the equations of motion
$\ddot X - X''=0$ and the closed string boundary condition 
$X^\mu(\sigma+\pi)=X^\mu(\sigma)$ 
we get the string modes expansion  which reads
\begin{eqnarray*}
X^\mu (z,\bar z)&=&
\frac{1}{2}\left( X_L^\mu (z)+ X_R^\mu (\bar z) \right)
\nonumber\\
X_L^\mu (z) 
&=&
x_{L}^\mu  -2\alpha' i p_{ L}^\mu  \ln(z) 
+i \sqrt{2\alpha'} \sum_{n\ne 0} \frac{sgn(n) a_n^\mu }{\sqrt{|n|}} z^{- n}
\nonumber\\
X_R^\mu (\bar z) 
&=&
x_{ R}^\mu  -2\alpha' i p_{ R}^\mu  \ln(\bar z) 
+i \sqrt{2\alpha'} \sum_{n\ne 0} \frac{ sgn(n) \tilde a_n^\mu }{\sqrt{|n}|} {\bar z}^{- n}
\end{eqnarray*}
where $z=e^{2i(\tau+ \sigma)}=e^{2(\tau_E+ i\sigma)}$ 
( $0 \le \sigma \le \pi$ ), 
$x^0_L=x^0_R$ and $p^0_L=p^0_R$.

The canonical momentum density is given by 
\begin{eqnarray*}
{\cal P}_\mu
&=&
\frac{1}{2\pi\alpha'} 
\left( G_{\mu\nu} \dot X^\nu -B_{\mu\nu} X'^\nu \right)
\end{eqnarray*}
so that from the canonical quantization we find
the non vanishing commutators
\begin{eqnarray*}
 ~[x^\mu_L, p^\nu_L ] &=& 
 ~[x^\mu_R, p^\nu_R ] = 
 i G^{\mu \nu}
\nonumber\\
 ~[ a^\mu_n, a^\nu_m ] &=& 
 ~[\tilde a^\mu_n, \tilde a^\nu_m ] = 
  G^{\mu \nu}\, sgn(n)\, \delta_{n+m,0}
\end{eqnarray*}
and the Hamiltonian density 
\begin{eqnarray*}
{\cal H} &=&
\frac{1}{4\pi\alpha'} G_{\mu \nu}\left(  \dot X^\mu \dot X^\nu + X'^\mu X'^\nu
\right)
\nonumber\\
&=&
\frac{1}{4\pi\alpha'} 
\left(
G^{\mu \nu} 
\left( 2\pi\alpha' {\cal P}_\mu +B_{\mu \lambda} X'^\lambda\right)
\left( 2\pi\alpha' {\cal P}_\nu +B_{\nu \kappa} X'^\kappa\right)
+ G_{\mu \nu} X'^\mu X'^\nu
\right)
\end{eqnarray*}

Using the previous quantization the energy momentum tensor 
\begin{eqnarray*}
T(z)
=-\frac{1}{4\alpha'}: \partial X^\mu_L G_{\mu \nu} \partial X^\nu_L:
= \sum_k \frac{L_{k  }}{z^{k+2}}
&\spazio&
\bar T(\bar z)
=-\frac{1}{4\alpha'}: \bar\partial X^\mu_R G_{\mu \nu}  \bar\partial X^\nu_R:
= \sum_k \frac{\bar L_{k  }}{\bar z^{k+2}}
\end{eqnarray*} 
with in particular 
\begin{eqnarray*}
L_0 
&=&
\alpha' p_L^\mu G_{\mu \nu} p_L^\nu 
+ \sum_{n=1}^\infty n a^{\mu\dagger}_n G_{\mu \nu} a_n^\nu
\nonumber\\
\tilde L_0 
&=&
\alpha' p_R^\mu G_{\mu \nu} p_R^\nu 
+ \sum_{n=1}^\infty n \tilde a^{\mu \dagger}_n G_{\mu \nu}
\tilde a_n^\nu
\end{eqnarray*}
implies that the vacuum is defined by
\begin{equation}
p_L^\mu|0,\tilde 0\rangle= p_R^\mu|0,\tilde 0\rangle
=a_n^\mu |0,\tilde 0\rangle = \tilde a_n^\mu |0,\tilde 0\rangle=0
\spazio
n> 0
\label{c_vacuum}
\end{equation}
and it is normalized as
\begin{eqnarray*}
\langle k_0, n ,w | k'_0, n', w'\rangle
&=&
2\pi \delta(k_0-k'_0) \delta_{n,n'} \delta_{w,w'}
\end{eqnarray*}
where we have defined 
$p^\mu_L|k_L\rangle=G^{\mu \nu}k_{L \nu}|k_L\rangle$  with 
$|k_L\rangle=e^{ik_{L \mu}  x_L^\mu}|0,\tilde 0\rangle 
$, 
$\langle k_L| =|k_L\rangle^\dagger$,
 similarly for the other momenta and the possible values of $k_L, k_R$
 together with the definitions of $n,w$
 are given in (\ref{p_L+p_R_spectrum}).

Because of the space periodicity the translation generator
\begin{eqnarray*}
T&=& ||T_i ||
= \int^\pi_0 d\sigma \, {\cal P} = E^T p_L +E p_R
\end{eqnarray*}
along with eq. (\ref{X_closed_periodicity}) implies
that the operators $p_L = || p^i_{ L} ||$
and $p_R = || p^i_{ R} ||$ have spectrum 
\begin{eqnarray}
p_L &=& G^{-1} k_L = \frac{1}{2} G^{-1} \left( n + E \frac{w}{\alpha'} \right)
\nonumber\\
p_R &=& G^{-1} k_R =\frac{1}{2} G^{-1} \left( n - E^T \frac{w}{\alpha'} \right)
\label{p_L+p_R_spectrum}
\end{eqnarray}
where $E=G+B$, $E^T=G-B$ ( with $G=|| G_{i j} ||$, $B=|| B_{i j} ||$ )
and
$n = || n_i || $, $w = || w^i || $ are  integer valuated vectors.

Following \cite{Giveon:1988tt} we consider T-duality as a canonical
 transformation given by\footnote{
For the time direction we have trivially
$${\cal P}_0=  {\cal P}_0^t
\spazio
 X^{ 0}= X^{t\, 0}  $$.
}
\begin{eqnarray}
{\cal P}_a
=  
\frac{1}{2\pi\alpha'}
\gamma_{a b} X^{t '}{}^{ b}
&\spazio&
{\cal P}_a^t
= 
\frac{1}{2\pi\alpha'}
 X'^{ b} \gamma_{b a}
\nonumber\\
{\cal P}_m
=  
{\cal P}_m^t
&\spazio&
 X^{ m}
= 
 X^{t\, m} 
\label{closed-T-duality}
\end{eqnarray}
where we have split $i,j,\dots$ 
in $a,b,\dots$ for the ``perpendicular'' directions 
which we T-dualize and $ m,n,\dots$ for the other ``parallel'' directions,
$X^{t\, i},\calP^t_j$ are the new T-dual coordinates,
and $|| \gamma_{a b} ||\in \alpha' SL(d_\perp,\Z)$ \footnote{
$\gamma_{a b} \in \alpha' \Z$ because we want that the new theory
  described by the new canonical variables be equivalent theory to the
  old one, this in particular means mapping integers $n$ and $w$ into
  new integers $n^t$ and $w^t$.
}is a constant matrix, 
usually taken proportional to the unity and $d_\perp$ is the number of
perpendicular directions.
These relations imply the usual transformations
\begin{eqnarray}
\partial X^t =  (P_\perp \gamma^{-1}E^T +P_\parallel) \partial X
&\spazio&
\bar \partial X^t =  - (P_\perp \gamma^{-1}E - P_\parallel) \bar \partial X
\label{Xt_X_relation}
\end{eqnarray}
where we have introduced the projector $P_\perp$ ($P_\parallel$) on
the directions we (do not) T-dualize and defined 
$\gamma_{m n}=\alpha' \delta_{m n}$, $\gamma_{a m}=\gamma_{m a}=0$. 
These relations can be extended to zero modes.
The matrix $E=G+B$ transforms then as
\begin{eqnarray*}
E^t
&=&
(P_\perp \gamma^{T} - P_\parallel E)(P_\perp \gamma^{-1}E -P_\parallel)^{-1}
\end{eqnarray*}
The inverse relations can be trivially obtained exchanging $\gamma
\leftrightarrow \gamma^T$ as follows from
eq. (\ref{closed-T-duality}), i.e. for example
\begin{eqnarray}
E
&=&
(P_\perp \gamma - P_\parallel E^t)(P_\perp \gamma^{-T}E^t -P_\parallel)^{-1}
\label{E_Et}
\end{eqnarray}

\COMMENTO{
\begin{eqnarray}
G^t
&=&
\gamma^T E^{-1} G E^{-T} \gamma
=
\gamma^T E^{-T} G E^{-1} \gamma
\label{Gt}
\\
B^t
&=&
-\gamma^T E^{-1} B E^{-T} \gamma
=
-\gamma^T E^{-T} B E^{-1} \gamma
\label{Bt}
\end{eqnarray}
}


\subsection{The ``neutral'' string.}

The action for a ``neutral'' string in constant open and closed background
can be written as
\begin{eqnarray*}
S
&=&
\int d\tau\, \int^\pi_0 d\sigma 
\left(
G_{\mu\nu}\frac{ \dot X^\mu \dot X^\nu - X'^\mu X'^\nu}{4\pi\alpha'}
-\calF_{\mu\nu} \dot X^\mu  X'^\nu 
\right)
\nonumber\\
&&-\int  d\tau\, 
\left(
e_{(0)} a_{(0)\mu} \dot X^\mu|_{\sigma=0}
-e_{(\pi)} a_{(\pi)\mu} \dot X^\mu|_{\sigma=\pi}
\right)
\end{eqnarray*}
where $\calF = F+\frac{B}{2\pi\alpha'}$,
$e_{(0)}$ ($e_{(\pi)}$) is the charge at $\sigma=0$ ($\sigma=0$) and 
$a_{(0)\mu}$ ($a_{(\pi)\mu}$) is the constant gauge field felt
by the string at $\sigma=0$ ($\sigma=\pi$) and that cannot be gauged away
on a circle.
It is noteworthy to stress that the previous expression is valid when
$F=e_{(0)} F_{(0)}= e_{(\pi)} F_{(\pi)}$ ( $F_{(0)}$  being the field
strength at $\sigma=0$ and similarly for $F_{(\pi)}$ ) 
and does not imply that the
open string ends on the same brane and has the same charges (even if
charges can only be $\pm 1$) 
but only the degeneracy of the
force felt by the two string endpoints: this is the reason of the quotes
around neutral in the title.
We assume moreover that the two branes\footnote{
When $e_{(0)}=e_{(\pi)}$ and $a_{(0)}=a_{(\pi)}$ the open string could
actually have both endpoints on the same brane and therefore the
system could be made of just one brane. This is a special case but all
what follows goes through in the same way up to trivial modifications. 
} where the string ends have non
trivial but equal homology (similar and more general 
configurations are considered in \cite{Bianchi:2005yz}) described by
\COMMENTO{
\begin{eqnarray}
X^i(\sigma^0,\sigma+2\pi s) = X^i +2\pi W^i
\spazio 
W^i\in N^*
\label{homology}
\end{eqnarray}
}
\begin{eqnarray}
X(\sigma^0,\sigma^i+2\pi s^i)= X(\sigma^0,\sigma^i) +2\pi Ws
\spazio 
\forall s\in Z^d
\,\,
W^i_{.\, i}\in N^{*}
\label{homology}
\end{eqnarray}
where $\sigma$ are the worldvolume coordinates of the D25 and we have
fixed $0\le \sigma^i < 2\pi$.
Notice that we write this and all following expressions as if we used 
a generic matrix $W$ even if we actually use $W=diag(W^1,W^2,\dots,W^d)$.
By this we mean that the fields living  on the branes must have that
periodicity, for example in the case of tachyons this means 
$T_{(0)}(x^0,x^i)=T_{(0)}(x^0,x^i+2\pi W^i_j s^j)$ with $x=X(0,\tau)$,
$T_{(\pi)}(x^0,x^i)=T_{(\pi)}(x^0,x^i+2\pi W^i_j s^j)$ with $x=X(\pi,\tau)$.
More general situations are possible where the homology of the two
branes  are different but we will not discuss them here since they can
be treated efficiently with the boundary state formalism.

The  boundary conditions are now
\begin{eqnarray}
&& G_{\mu \nu} X'^\nu
+ 2\pi\alpha'\, \dot X^\nu \calF_{\nu\mu} |_{\sigma=0,\pi}= 0
\label{Neu tral_Xopen_b c}
\end{eqnarray}
which implies the modes expansion
\begin{eqnarray}
X^\mu(\sigma,\tau) 
&=&
x^\mu
+2\alpha' (\delta^\mu_\nu \tau 
+ 2\pi\alpha'\,G^{\mu\kappa} \calF_{\kappa \nu}\,\sigma) p^\nu 
\nonumber\\
&&+i \sqrt{2\alpha'} \sum_{n\ne 0} \frac{sgn(n) e^{-i n\tau}}{\sqrt{|n|}}
 \left(\delta^\mu_\nu \cos(n\sigma) 
- i 2\pi\alpha'\,G^{\mu\kappa} \calF_{\kappa \nu} \sin(n\sigma) \right) a^\nu_n
\label{Xopen_modes_expansion}
\nonumber\\
~~~~
\end{eqnarray}

The non vanishing commutation relations \cite{Abouelsaood:1986gd,Chu:1999gi} are
\begin{eqnarray*}
~[x^0, p^0 ] =i \calG^{0 0}
&\spazio&
~[a^0_m, a^0_n ]
=
\calG^{0 0}\, sgn(m)\, \delta_{m+n,0}
\nonumber\\
~[x^i, p^j ] =i \calG^{i j}
&\spazio&
~[x^i, x^j ] = i \theta^{i j}
\nonumber\\
~[a^i_m, a^j_n ]
&=&
\calG^{i j}\, sgn(m)\, \delta_{m+n,0}
\end{eqnarray*}
where we have defined 
$\calE=||\calE_{i j} ||= G+2\pi\alpha' \calF=
E+2\pi\alpha' F$, 
the open string metric \cite{Seiberg:1999vs} as
\begin{eqnarray}
&&\calG_{0 0}=G_{0 0}
\spazio
 \calG_{0 i}=0
\nonumber\\
&&\calG=||\calG_{i j}||=G-(2\pi\alpha')^2\,\calF G^{-1}\calF=\calE^{T} G^{-1} \calE
\label{calG}
\end{eqnarray}
and $\calG^{-1}=\calE^{-T} G \calE^{-1}=\calE^{-1} G \calE^{-T}$,
$\theta=(2\pi\alpha')^2\, \calE^{-T} \calF \calE^{-1}$.

Using the previous quantization the energy momentum tensor 
\begin{eqnarray*}
T(z)
&=&-\frac{1}{\alpha'}: \partial X^\mu G_{\mu \nu} \partial X^\nu:
= \sum_k \frac{L_{k  }}{z^{k+2}}
\end{eqnarray*} 
with in particular
\begin{eqnarray}
L_0 
&=&
\alpha' p^\mu \calG_{\mu\nu} p^\nu 
+ \sum_{n=1}^\infty n a^{\mu\dagger}_n \calG_{\mu\nu} a_n^\nu
\label{L0_open}
\end{eqnarray}
implies that the vacuum is defined by
\begin{equation}
p^\mu|0\rangle
=a_n^\mu |0\rangle = 0
\spazio
n> 0
\label{o_vacuum}
\end{equation}
and it is normalized as%
\COMMENTO{
\footnote{
CHECK!!
We use also 
\begin{eqnarray*}
<x|n>=\frac{ e^{i k_i x^i} }{(2 \pi)^{d/2} (\det(W G W) )^{1/4} }
&\spazio&
1= \int d^d x\, |x\rangle \sqrt{ \det(W G W) } \langle x| 
\end{eqnarray*}
with $0\le x^i \le 2\pi W^i$.
}
}
\begin{eqnarray*}
\langle k_0,n | k'_0,n'\rangle
&=&
2\pi \delta(k_0-k'_0) \delta_{n, n'}
\end{eqnarray*}
where we have defined $p^\mu |k \rangle=\calG^{\mu \nu}k_{ \nu}
|k\rangle$  with 
$|k\rangle = | n \rangle =e^{i k_{ \mu}  x^\mu}|0\rangle 
$, 
$\langle k| =|k\rangle^\dagger$
and the relation between $k$ and $n$ is given by eq. (\ref{p_spectrum}).

Since we have assumed the space be compact from the expression for the space
translation generators
\begin{eqnarray*}
T
&=&
||T_i||=
\int_0^\pi d\sigma\, {\cal P}
=- (e_{(0)} a_{(0)}- e_{(\pi)} a_{(\pi)}) +
\calG p
\end{eqnarray*}
where we have used 
$\calG=G-(2\pi\alpha')^2\,\calF G^{-1}\calF=\calE^{T} G^{-1} \calE$,
we get the spectrum for the $p$ operator to be\footnote{
We write $W^{-T}$ and not $W^{-1}$ even if the $W$ matrix we consider
now is diagonal since this is the proper way of writing the momenta in
presence of non diagonal $W$s.
}
\begin{eqnarray}
p&=&\calG^{-1}k 
=\calG^{-1}( W^{-T}n+(e_{(0)} a_{(0)}- e_{(\pi)} a_{(\pi)}) )
\label{p_spectrum}
\end{eqnarray}
where $W=diag(W^1,W^2,\dots,W^d)$ and 
$ 0\le \left( e_{(0)} a_{(0) i}- e_{(\pi)} a_{(\pi) i} \right) <
\frac{1}{W^i}  $ \footnote{
In the case where $W$ were not diagonal, from the gauge invariance of
the zero modes part of the ``generalized'' velocity matrix elements 
$\langle\Psi| \frac{1}{2\pi\alpha'} 
\left( G_{\mu\nu} \dot X^\nu -B_{\mu\nu} X'^\nu \right) |\Psi'\rangle$
we get 
$W^T (e_{(0)} a_{(0)}- e_{(\pi)} a_{(\pi)}) \equiv
W^T (e_{(0)} a_{(0)}- e_{(\pi)} a_{(\pi)}) +m
\spazio
\forall m\Z^d
$.
}. 

All the field strengths are quantized as
follows from their first Chern class\footnote{
We have used $F_{\mu \nu}=\partial_\mu A_\nu -\partial_\nu A_\mu $.
} 
\begin{eqnarray*}
c_1=\frac{1}{2\pi}\int_{\Sigma^{\bar i \bar j}} F
= 2\pi  \left( W^T  F W \right)_{\bar i \bar j} 
=2 \pi F_{\bar i \bar j} W^{\bar i} W^{\bar j}
\in \Z
\spazio \forall \bar i,\bar j
\label{Chern_1}
\end{eqnarray*}
since the space is compact and where we have defined the cycle 
$\Sigma^{\bar  i \bar j}=
\{ X^i= 2\pi\left( W^i_{\bar i} t^{\bar i}+W^i_{\bar j} t^{\bar
  j}\right)\, , 0\le t^{\bar i}, t^{\bar j}< 1\}$.

What we have written until now is valid for a $D25$ brane.
In analogy to the superstring case where we actually can measure the
various branes charges, we can assert that
when all possible $F_{i j}$ are turned on we generically describe a
complex system of a $D25$ brane together 
with all possible lower dimensional branes.
While lower dimensional branes at angles can be obtained by T-duality and by
switching off some $F$ components, 
lower dimensional branes wrapping straight can formally obtained letting
some $F$ component go to infinity.
In a more explicit way we start by
splitting the space time indexes $\mu,\nu,\dots$ as $m,n,\dots$ for
directions parallel to the brane ($i,j,\dots$ for the spacial ones) 
and $a,b,\dots$ for directions
perpendicular to the brane, we set 
$F_{a m}=0$ to simplify and $F_{a b}=0$\, \footnote{
In sigma model under T-duality we roughly have 
$A_a(X^\mu) \dot X^a|_{\sigma=0} \Rightarrow \frac{1}{2\pi\alpha'}
\Phi_a(X^a,X^m) X^{'a}|_{\sigma=0} $  but equations of motions imply 
the constraints $\Phi_a(X^a,X^m)=0$ and hence we set $F_{a m}=F_{a b}=0$. 
%
} 
in order to avoid higher dimensional branes in the system
\footnote{
It is also natural to assume $W^a=1$ but 
this is again not strictly necessary since $W^a>1$ would give special 
cases where $W^a$ $Dp$
branes are located at regular interval in direction $x^a$ as it can be
seen by the momentum quantization.
%
}. Then boundary conditions read
\begin{eqnarray}
&& G_{m n} X'^n + G_{m a} X'^a
-  2\pi\alpha'\,  \calF_{m n}  \dot X^n  
|_{\sigma=0,\pi}= 0
\nonumber\\
&&X^a-x^a_0|_{\sigma=0}= X^a-x^a_\pi|_{\sigma=\pi}= 0
\label{Neutral_Xopen_b c_Dp}
\end{eqnarray}
which can formally obtained from eq. (\ref{Neu tral_Xopen_b c}) 
letting $F_{a b}\rightarrow\infty$ (if the number of
transverse directions is bigger than one) and noticing that the term 
$-  2\pi\alpha'\,  \calF_{m a}  \dot X^a  |_{\sigma=0,\pi}$ is then
identically zero because $\dot X^a|_{\sigma=0,\pi}=0$.

\section{Boundary states.}
\subsection{Boundary states for $D25$}
\label{sect_D25}
In order not to kludge the notation we start with the simplest case,
i.e. the $D25$ with non trivial homology given by eq. (\ref{homology}).
We can now rewrite the open boundary condition 
(\ref{Neu tral_Xopen_b c})
for the closed string as
\begin{eqnarray}
\left( G_{\mu \nu} \dot X^\nu
+ 2\pi\alpha'\,  X^{' \nu} \calF_{\nu\mu} \right) |_{\tau=0}
|D25(E,F,W)\rangle
&=& 0
\label{def_boundary_0}
\end{eqnarray}
This can also be reexpressed using the closed string momentum which
already incorporate all $B$ as
\begin{eqnarray}
\left( \calP_\mu - F_{\mu\nu} X^{' \nu} \right)   |_{\tau=0}
|D25(E,F,W)\rangle
&=& 0
\label{def_boundary}
\end{eqnarray}
This expression is already interesting since it reveals that under
T-duality $B$ does not play the same role of  $F$.

The previous expression can be analyzed in modes to give
\begin{eqnarray*}
\left( \calE^T \alpha_n -\calE \tilde \alpha_{-n} \right) |D25(E,F,W)\rangle
&=& 0
\end{eqnarray*}
In particular the zero mode sector yields
\begin{eqnarray}
\left( \hat n -2\pi F \hat w \right) |D25(E,F,W)\rangle = 0
\label{B_zero_modes}
\end{eqnarray}
where $\hat n$ and $\hat w$ are the operators, obtained by a linear
combination of $p_L$ and $p_R$,  which have the corresponding unhatted 
quantities defined in eq. (\ref{p_L+p_R_spectrum})  as
eigenvalues.
These expressions imply the possibility of rewriting the spectrum 
of $p_L$, $p_R$
as a function of $w$ only as
\begin{eqnarray*}
p_L=\frac{1}{2}G^{-1}\calE \frac{w}{\alpha'}
&\spazio&
p_R=-\frac{1}{2}G^{-1}\calE^T \frac{w}{\alpha'}
\end{eqnarray*}

It is now immediate to deduce that all the entries of the matrix
$2\pi F=||2\pi F_{i j }||$ must be rational otherwise the previous
constraint would not have any solution.

Actually it turns out that also $\hat n$ must have { \sl rational }
eigenvalues when acting on the boundary and not only integers as it
comes from the closed string spectrum.
To explore this point let us consider the simplest case where only
$F_{1 2} \ne 0$, explicitly $2\pi F_{1 2} = \frac{p}{q}$ ($p,q\in\Z$)  
then the non trivial equations are
\begin{eqnarray}
\left( \hat n_1 - \frac{p}{q} \hat w^2 \right) |D25(E,F,W)\rangle &=& 0
\nonumber\\
\left( \hat n_2 + \frac{p}{q} \hat w^1  \right) |D25(E,F,W)\rangle &=& 0
\label{simplest_B_1}
\end{eqnarray}
If we insist on having only integer $n$ then we must conclude that
$(w^1,w^2)=q (l^1,l^2)$ and $(n_1,n_2)=p (l^2,-l^1)$ with arbitrary
integers $l^1$ and $l^2$. We can now take the simplest boundary given
by the sum of all allowed $|n,w\rangle$ with equal coefficients, i.e.
the zero modes non trivial part of the boundary is given by
$|D25(E,F,W)_{(0)}\rangle \sim \sum_{(l^1,l^2)\in\Z^2} |n=p (l^2,-l^1),w=q
(l^1,l^2) \rangle$.
If we now compute the boundary-boundary interaction as $\langle D25(E,F,W)
| e^{-t (L_0 +\tilde L_0)} |D25(E,F,W)\rangle$ we see immediately that we
get a contribution from these zero modes of the form
$\sum_{(l^1,l^2)} e^{ -{t\over 2} q^2 (l^1,l^2)  {\calG_2\over \alpha'} (l^1,l^2)^T }$ 
($\calG_2 $ being the $\calG$ submatrix with indexes running on 1 and
2 only).
Performing a Poisson resummation\footnote{
The Poisson formula is as follows
\begin{eqnarray*}
\sum_{n\in\Z^d} e^{-\frac{1}{2} n^T A n + i n^T B }
&=&
\frac{1}{\sqrt{\det A}} \sum_{m\in\Z^d} e^{-\frac{1}{2} (2\pi m +B )^T
  A^{-1} (2\pi m + B) }
\end{eqnarray*}
} on the previous expression we find
something proportional to
$\sum_{(m_1,m_2)\in\Z^2} 
e^{ -{2\pi^2 \over t} \alpha' {1\over q}(m_1,m_2)  \calG_2^{-1} 
{1\over q}(m_1,m_2)^T }$
which we want to interpret in the open channel.
We are therefore led to take the open string momenta of the form
$\calG_2^{-1}{1\over q}(m_1,m_2)^T$. From the $F$ quantization condition 
(\ref{Chern_1}) we
know that $q=W^1 W^2$ and therefore we find a {\sl wrong} open spectrum
which should be of the form $\calG_2^{-1}({n_1 \over W^1},{n_1 \over
  W^2})$ 
as in eq. (\ref{p_spectrum}).

We notice that the problem becomes even worst when we consider the
case where more $2\pi F_{i j}\propto \frac{1}{W^i W^j}$ are turned on,
since in this case in the generic situation $q\propto \prod W^i$
(where the product is over all directions where there is anon
vanishing $F$) in order to have integer $n$s.

Let us now assume that the $(n_1,n_2)=({l^2 \over W^1},-{l^1 \over W^2})$
then the possible winding are now $(w^1,w^2)= (W^1 l^1,W^2 l^2)$ again
with arbitrary integers $l^1$ and $l^2$. Performing the same steps as
before we find that the closed contribution is now
$\sum_{(l^1,l^2)} e^{ -{t\over 2}  (W^1 l^1,W^2 l^2)  {\calG_2\over
    \alpha'} (W^1 l^1, W^2 l^2)^T }$ which gives the right expression
when Poisson resummed, i.e. a contribution which is proportional to  
$\sum_{(m_1,m_2)\in\Z^2} 
e^{ -{2\pi^2 \over t} \alpha' (m_1,m_2) W^{-1} \calG_2^{-1} 
W^{-1}(m_1,m_2)^T }$

This is not completely unexpected. 
In the ``trivial'' case
 where $F=0$ and branes still have a non trivial homology the zero modes
 contribution to open string partition function is 
$\sum_{(m_1,m_2)\in\Z^2} 
e^{ -\tau \alpha' (m_1,m_2) W^{-1} G_2^{-1} W^{-1}(m_1,m_2)^T }$
which can be reproduced in the closed string channel if we take a
 boundary state  $|D25(E,F,W)_{(0)}\rangle 
\sim \sum_{(l^1,l^2)\in\Z^2} |n=0,w=(W^1 l^1,W^2 l^2) \rangle
= \sum_{(l^1,l^2)\in\Z^2} |n=0,w=W ( l^1,l^2) \rangle$, i.e. summing over some (not all!) of the
 possible closed string states which satisfy (\ref{B_zero_modes}) for $F=0$.
But now consider the case of boundary states which describe
open string interactions on parallel branes without background fields
 \cite{Pesando:2003ww}\footnote{
The results in \cite{Pesando:2003ww} are actually obtained in the case
 of trivial homology but can easily extended to more complex cases
 where the open string momenta are rational 
 since all results are obtained when arbitrary Wilson lines turned on.
} then we know that these boundaries are
produced by acting on the usual boundaries by the open string vertex
operators where the open string fields are substituted by the left
moving closed string fields. In presence of a brane winding $W^1, W^2$
times in $x^1,x^2$ directions we expect that the exponential part of
the open string vertexes being of the form $V\sim e^{i({m_1 \over
    W^1},{m_1 \over W^2}) ( X^1, X^2)^T } $ and this explains why we can
find non integer $n$\footnote{
There is actually a further subtlety since we have asserted that  the
open string vertexes are made using the left moving part of closed
fields only. As shown in \cite{Pesando:2003ww} there is actually a further
contribution to these interacting boundaries whose aim is to  divide
 equally the open string momenta in the left and  right moving sector
 and this fixes completely all the details.
}. We can reformulate this by saying that these non integer momenta are
 due to the fact that open string sees directions of length $2\pi W^i$
 while closed string sees a length of $2\pi$ and therefore we need
 adding more (non physical) states to closed string in order to
 describe open string states which explore shorter distances.
 
Because of this interpretation of these extra states it must be
stressed that they do not represent new physical closed string
excitations, i.e. they do not belong to the physical closed string
Hilbert space but they are the closed string representations of open
string physical states. Nevertheless they do propagate in space time
even if the do not interact with closed string while away from the brane.

With this understanding we can now write the complete boundary state
as
\begin{eqnarray}
|D25(E,F,W,y)\rangle
&=& {T_d(E,F)\over 2}
|D25(E,F)\rangle_{time}\,
|D25(E,F,W,y)\rangle_{space}\,
|D\rangle_{ghost}
\nonumber\\
|D25(E,F)\rangle_{time}\,
&=&
e^{-\sum_{n=1}^\infty  a^{\dagger 0}_n G_{0 0}  \tilde a^{\dagger 0}_n}
|k^0=0>
\nonumber\\
|D25(E,F,W,y)\rangle_{space}
&=&
e^{-\sum_{n=1}^\infty  a^\dagger_n G \calE^{-T} \calE \tilde
  a^{\dagger T}_n}
\sum_{w\in I(F,W) }
c_w(y) | n=2\pi F w,w>
\nonumber\\
|D\rangle_{ghost}\,
&=&
e^{\sum_{n=1}^\infty c_{-n}\tilde{b}_{-n}+\tilde{c}_{-n}b_{-n}
}
\frac{c_0+\tilde{c}_0}{2}
| q=\tilde q=1\rangle
\label{D25}
\end{eqnarray}
where $I(F,W)$ is the lattice given by 
$I(F,W)=\{w\in\Z^d | \forall i \spazio\&
\spazio w^i \propto W^i  \spazio\&
\spazio n=2\pi F w\}$ \footnote{
Since $2\pi F_{i j}={f_{i j}\over W^i W^j}$ with $f_{i j}\in\Z$ and
$w^i=W^i s^i$ with $s^i\in\Z$ this
means that $ n_i={f_{i j} s^j\over W^i}\propto{1\over W^i}$.
},
$c_w(y)$ are in principle arbitrary complex numbers but because of the
open string channel interpretation they turn out to be 
$c_w(y)=e^{2\pi i  w^T y}$  with 
$y_i=e_{(0)} a_{(0) i}$,
the tension 
$T_d(E,F)
$ can be obtained as in
\cite{DiVecchia:1997pr,DiVecchia:1999rh} 
and in \cite{Murakami:2002yd}
We have also introduced the usual ghost sector with energy momentum tensor
using the conventions of \cite{DiVecchia:1999rh} as
\begin{eqnarray*}
T^{(b c)}
&=&
-2 b\partial c - \partial b c
=\sum \frac{L_n^{(b c)}}{z^{n+2}}
\nonumber\\
L_0^{b c}&=&\sum_{n=0}^\infty n \left(b^\dagger_n c_n - c^\dagger_n b_n\right)
\end{eqnarray*}
similarly for the right moving sector 
and defined the $|q=1 \tilde q=1\rangle$ vacuum from the $SL(2,\C)$
invariant ghost vacuum  as $|q= \tilde q=1\rangle = c_1 \tilde
c_1|q= \tilde q=0\rangle $  such as $c_{n}|q=\tilde q=1\rangle=\tilde
c_{n}|q=\tilde q=1\rangle=0$ for $n \ge 1$.

\subsection{Modular transformation}
We can now compute the boundary-boundary interaction as usual
as
\begin{eqnarray*}
&&\langle D25(E,F_1,W_1,y_1)| e^{-t(L_0^{(X+b c)} +\tilde
    L_0^{(X+b c)})}
(b_0+\tilde b_0)(c_0-\tilde c_0) |D25(E,F_2,W_2,y_2)\rangle
=
\nonumber\\
&&\left({T_d\over 2} \right)^2 V_{time}
e^{2t}\left(  \prod_{n=1}^\infty (1-e^{-2 n t}) \right)^{2-1}
\left( \prod_{n=1}^\infty  \det(1-\calE_1^{-1} \calE_1^T \calE_2^{-T}
\calE_2 e^{-2n t} ) \right)^{-1}
\nonumber\\
&&\sum_{w\in I(F_1,W_1)\cap I(F_2,W_2)}
c^*_w(y_1) c_w(y_2)
\delta_{(F_1-F_2)w,0}
e^{-\frac{t}{2} w^T {\calG \over \alpha'} w}
\end{eqnarray*}
where we have inserted $(b_0+\tilde b_0)(c_0-\tilde c_0)$ to account
for the surviving part of the $SL(2,\C)$ symmetry.
The complete boundary-boundary amplitude reads\footnote{
We use as in \cite{DiVecchia:1999rh} the closed string propagator to
be $D=\frac{\alpha'}{2} \delta_{L_0-\tilde L_0,0} \int_0^\infty dt\, 
e^{-t(L_0^{(X+b c)} +\tilde L_0^{(X+b c)})} $.
}
\begin{eqnarray*}
&&{\cal A}(E,F_1,W_1,y_1,F_2,W_2,y_2)
=
\nonumber\\
&&\,\,\,
\frac{\alpha'}{2}
\int_0^\infty dt\,\langle D25(E,F_1,W_1,y_1)| 
e^{-t(L_0^{(X+b c)} +\tilde L_0^{(X+b c)})} |D25(E,F_2,W_2,y_2)\rangle 
\end{eqnarray*}
which in special case where $F=F_1=F_2$  and $W_1=W_2$ becomes
\begin{eqnarray*}
{\cal A}(E,F,W,y_1,y_2)
&=&
\frac{\alpha'}{2}
\left({T_d\over 2} \right)^2 V_{time}
\int_0^\infty dt\,
\left(  \prod_{n=1}^\infty  (1-e^{-2 n t}) \right)^{2-1-d}
\nonumber\\
&&\sum_{w\in I(F,W)}
c^*_w(y_1) c_w(y_2)
e^{-\frac{t}{2} w^T {\calG \over \alpha'} w}
\end{eqnarray*}
which can be transformed by a modular transformation of parameter
$\tau=\frac{2\pi^2}{t}$ into the corresponding open channel annulus
free energy when $D=26$
\begin{eqnarray*}
F(E,F,W,a_{(0)}-a_{(\pi)} )
&=&
2\frac{1}{V_{time} \sqrt{\det W \calG W} }
\int^\infty_0 {d\tau\over 2\tau}\, Tr'(e^{-\tau L_0^{(X+b c)}})
\end{eqnarray*}
where $Tr'$ means a trace over all oscillators but ghost zero modes
and the factor $V_{time} \sqrt{\det W \calG W}$ is inserted for
allowing the decompactification limit,
the trace is over all modes but the ghost zero modes.

\section{T-duality and boundary states for $Dp$ branes.}
We want now to discuss the action of T-duality on the boundary state.
We start our discussion from eq. (\ref{def_boundary}) and not from
eq. (\ref{def_boundary_0}) because T-duality is better defined at the
canonical level by eq.s (\ref{closed-T-duality}) in a $B$ independent way.
Applying the T-duality transformation on  eq. (\ref{def_boundary})
written in term of $X^t$ and $\calP^t$
we get
\begin{eqnarray}
\calP_m - \Ft_{m n} X^{' n} -2\pi\alpha' \Ft_{m a} (\gamma^{-1})^{a b}
\calP_b |_{\tau=0}
|D25(E^t,\Ft,W^t,y^t)\rangle
&=& 0
\nonumber\\
\frac{1}{2\pi\alpha'}(\gamma^{T})_{a b} X^{' b} 
- \Ft_{a m} X^{' m} -2\pi\alpha' \Ft_{a b} (\gamma^{-1})^{b c} \calP_c   |_{\tau=0}
|D25(E^t,\Ft,W^t,y^t)\rangle
&=& 0
\nonumber\\~
\label{T-duality-def_boundary}
\end{eqnarray}
where $a,b ,\dots$ run on the $d_\perp=d-p$ directions ``perpendicular'' to the brane,
i.e. on the directions along which we T-dualize,  and
$m,n,\dots $ on the $p+1$ directions parallel to the brane.
The second equation clearly shows that the geometrical embedding properties
(the ``angles'' of the brane ) are only determined by $\Ft_{a m}$ when $\Ft_{a
  b}=0$ (the would be field strengths in the directions transverse to
the brane which describe higher dimensional branes than our $Dp$ ,
i.e. we describe a system without higher dimensional branes than
our $Dp$ ); 
in fact eq. (\ref{T-duality-def_boundary}) can be
rewritten as
\begin{eqnarray*}
\left[  X^{ a} - (\alpha' \gamma^{-T})^{a b} 2\pi \Ft_{b m} X^{ m} \right]' |_{\tau=0}
|D25(E^t,\Ft,W^t,y^t)\rangle
&=& 0
\label{Geometrical_embedding}
\end{eqnarray*}
which is interpreted in the open channel as the constraint that one
open string endpoint must 
be  on the hyperplane $X^{ a} - (\alpha' \gamma^{-T})^{a b} 2\pi \Ft_{b
  m} X^{ m}=const$ which, as it should, has rational ``angles'' $(\alpha'
\gamma^{-T})^{a b} 2\pi \Ft_{b m}$.

This is actually a special case which can be obtained as the (careful)
 limit 
 $\Ft_{a b}\rightarrow 0$ of 
the generic case with $\det( \Ft_{a b})\ne 0$ \,\footnote{
This requires that we T-dualize at least 2 directions.
} in which case
eq.s (\ref{T-duality-def_boundary}) can be recast in the same form
of eq. (\ref{def_boundary}) as
\begin{eqnarray*}
\left( \calP_\mu - F_{\mu\nu} X^{' \nu} \right)   |_{\tau=0}
|D25(E^t,\Ft,W^t,y^t)\rangle
&=& 0
\end{eqnarray*}
where now the field strength is given by
\begin{eqnarray}
||F_{\mu \nu} ||
&=&
\left(
\begin{array}{cc}
\Ft_{m n} -\Ft_{m c} (F_{\perp \perp}^{t\, -1})^{c d} \Ft_{d n} 
&
\frac{1}{2\pi\alpha'} \Ft_{m c} (F_{\perp \perp}^{t\, -1})^{c d} (\gamma^T)_{d b}
\\
-\frac{1}{2\pi\alpha'} \gamma_{a c} (F_{\perp \perp}^{t\, -1})^{c d} \Ft_{d n}
&
\frac{1}{(2\pi\alpha')^2} \gamma_{a c} (F_{\perp \perp}^{t\, -1})^{c d} (\gamma^T)_{d b}
\end{array}
\right)
\label{F_Dp}
\end{eqnarray}
where $|| (F_{\perp \perp}^{t\, -1})^{c d} ||$ is the inverse of the matrix 
$F_{\perp \perp}^{t}=|| F^{t}_{a b}||$.
When we express the boundary $|D25(E^t,\Ft,W^t,y^t)\rangle$ using the
operators $a,\tilde a$ and the matrices $E, F$ we get
\begin{eqnarray}
|Dp(E,F,W,y)\rangle
&=& 
|D25(E^t,F^t,W^t,y^t)\rangle
\nonumber\\
&=&
{T_d\over 2}
|D25(E^t,F^t)\rangle_{time}\,
|D25(E^t,F^t,W^t,y^t)\rangle_{space}\,
|D\rangle_{ghost}
\nonumber\\
 ~\label{Dp}
\\
|D25(E^t,F^t,W^t,y^t)\rangle_{space}
&=&
e^{-\sum_{n=1}^\infty  a^\dagger_n G \calE^{-T} \calE \tilde
  a^{\dagger T}_n}
\nonumber\\
&&\sum_{w^t\in I(\Ft,W^t) }
c_{w^t}(y^t) 
| n=2\pi F w,w>
\label{Dp0}
\end{eqnarray}
where the pieces not explicitly written are the same as in eq. (\ref{D25}).
To get this result we have used  
$n_a=\frac{\gamma_{a b}}{\alpha'}w^{ t b}$, 
$w^a=\alpha' (\gamma^{-T})^{a b} n^{ t}_{ b}$.
This has a somewhat dramatic consequence: now all $n_a$ are 
{\sl integer} while $w^a$ are {\sl fractional}, moreover $W^{t a}$ are
changed into a matrix $||W^{a b}||$ 
as discussed below, see eq. (\ref{twisted_W}) and $y$ are changed too
as in eq. (\ref{new_y})\footnote{
Notice that we have still written $c_{w^t}(y^t)=e^{2\pi i w^{t\, T}y^t}$
which must be reexpressed using the $Dp$ quantities $y$, $w$ and $n$
as
$c_{w}(y)=c_{w^t}(y^t)=e^{2\pi i (w^m y_m + n_a y^a)}$.
}.

\subsection{The non degenerate case: $\det( \Ft_{a b})\ne 0$}
To understand what is going on we consider again the simplest case we
considered in section \ref{sect_D25}, i.e. we take
$2\pi F_{1 2}= \frac{f_1 f_2}{W^1 W^2}=\frac{p}{q}$.
We rewrite the non trivial boundary equations (\ref{simplest_B_1}) 
for the involved zero modes as
\begin{eqnarray}
\left( \hat n_1 -  \frac{f_1 f_2}{W^1 W^2}\hat w^2 \right) |B\rangle &=& 0
\nonumber\\
\left( \hat n_2 +  \frac{f_1 f_2}{W^1 W^2} \hat w^1  \right) |B\rangle &=& 0
\label{simplest_B_2}
\end{eqnarray}
These equations have two different kinds of solution assuming that
either $n_1$ or $w^1$ be integer and similarly for $n_2$ and $w^2$\,
\footnote{
There is another solution one would think about:
\begin{eqnarray*}
\left\{\begin{array}{cc}
w^1= W^1 l^1 & n_2 = - \frac{f_1 f_2}{W^2} l^1 \\
n_1= f_1 j_1 & w^2 = \frac{W_1 W_2}{f_2} j_1 \\
\end{array}
\right.
l^1,j_1\in\Z
\end{eqnarray*}
but now neither $n_2$ nor $w^2$ are integers.
Moreover this solution is not the T-dual of the first solution.
}:
\begin{eqnarray}
``D2'':\,
\left\{\begin{array}{cc}
w^1= W^1 l^1 & n_1 = \frac{f_1 f_2}{W^1} l^2 \\
w^2= W^2 l^2 & n_2 = -\frac{f_1 f_2}{W^2} l^1 \\
\end{array}
\right.
l^1,l^2\in\Z
\label{D2}
\\
``D0'':
\left\{\begin{array}{cc}
n_1= f_1 j_1 & w^1 = \frac{W_1 W_2}{f_1} j_2 \\
n_2= f_2 j_2 & w^2 = -\frac{W_1 W_2}{f_2} j_1 \\
\end{array}
\right.
j_1,j_2\in\Z
\label{D0}
\end{eqnarray}
As we discuss below these two solutions have a clear different
meaning: an integer $w$ means a wound worldvolume direction while a
fractional $w$ and an
integer $n$ means a direction perpendicular to the worldvolume.
It is important also to realizes that given the defining equations 
for a boundary state (\ref{simplest_B_2})
there can be  many different solutions according to the interpretation we
give to the defining equations, besides the difference between integer
and rational $w$ we stated above 
there is a further ambiguity  associated with a given
$2\pi F_{1 2}= \frac{p}{q}$ since there can be many different ways of
factorizing $p=f_1 f_2$ and $q=W^1 W^2$ and each of this corresponds to
a different configuration.

The first solution given by eq.s (\ref{D2}) is the one already discussed in
section \ref{sect_D25} and describes two branes wrapped $W^1$ ($W^2$)
times along $X^1$ ($X^2$) %
\COMMENTO{
with a U(1) gauge bundle with 
first Chern Class $c_1=f_1 f_2$
}.

The  second solution eq.s (\ref{D0}) can be seen as the T-dual
solution of the previous where T-duality has been performed both along
$X^1$ and $X^2$ and it has fractional windings and integer momenta.
Given this viewpoint we can think that the branes worldvolumes are
orthogonal to the plane $X^1 X^2$.
Since the branes are at fixed position in the
coordinate $X^1$ and $X^2$  it is not unreasonable to interpret this
solution as due to the fact that the {\sl open} string sees a
fractional periodicity in these directions. From the values of momenta
in eq. (\ref{D0}) 
we see that the boundary is invariant for translations which are
multiple of $\frac{2\pi}{f_1}$ in direction $X^1$ and similarly for
$X^2$.
We suppose therefore that the open string has periodicity    
\begin{eqnarray*}
X^1_{(open)}\equiv X^1_{(open)}+ 2\pi\frac{N^1}{f_1}
&\spazio&
X^2_{(open)}\equiv X^2_{(open)}+ 2\pi\frac{N^2}{f_2}
\end{eqnarray*}
and, as we show below, it turns out that
\begin{eqnarray}
N^1&=& N^2 = W^1 W^2 .
\label{D0_periodicity}
\end{eqnarray}
\COMMENTO{
which implies that the would-be first Chern Class $c_1=\frac{N^1
  N^2}{W^1 W^2}$ is again integer.}
This would clearly be unacceptable if the branes wound along these
directions since fractional winding is meaningless.
Given this interpretation we can check as in section \ref{sect_D25}
that the open string zero modes contribution to the one loop free
energy which is proportional to 
$\sum_{(m_1,m_2)\in\Z^2} 
e^{ -\tau \alpha' (\frac{f_1}{N^1 }m_1,\frac{f_2}{N^2 }m_2)
  \calG_2^{-1} 
(\frac{f_1}{N^1} m_1,\frac{f_2}{N^2 } m_2)^T }$
is correctly reproduced after a Poisson resummation 
in the closed channel where the zero modes
give in the boundary-boundary interaction a contribution proportional to
$\sum_{(l^1,l^2)} e^{ -{t\over 2}  
 (\frac{W_1 W_2}{f_1} l^1,\frac{W_1 W_2}{f_2} l^2) {\calG_2\over \alpha'} 
(\frac{W_1 W_2}{f_1} l^1,\frac{W_1 W_2}{f_2} l^2)^T }$
only when eq. (\ref{D0_periodicity}) holds.

The open string sees a fractional periodicity in directions perpendicular
to the branes but the closed string sees an integer periodicity
therefore the picture from the closed string point of view is the
existence of $W^1 W^2$ {\sl identical} copies of the basic system of two branes
covering regularly the torus. Pictorially and naively 
the basic system of two branes can be portrayed
by  taking two
$Dp$ branes spaced along each direction $X^{a}$ $a=1,2$ by  
$2\pi\left( e_{(0)} a_{(0) a}- e_{(\pi)} a_{(\pi)a}\right)$,
inscribing them into a parallelogram with side lengths
$\frac{2\pi}{W^a}$  
and then tiling the $T^2$  torus with side lengths $2\pi$
using this configuration\footnote{
Since either $\frac{W_1 W_2}{f_1}$ or $\frac{W_1 W_2}{f_2}$ or both
can be greater then unity, we are actually tiling  a multiple of $T^2$.
}.
All systems in this configuration are identical because they
are the same system from the open string point of view.
This picture is the T-dual of the open string point of view of the
existence of closed strings which seem to be open when seen
from the multiple wound open strings perspective.

The existence of $W^1 W^2$ copies is also consistent with  the
computation of the
local energy density in this picture and its T-dual.
\COMMENTO{
The interpretation of this strange configuration is not so clear in
open string channel when $F\ne0$. When $F=0$ the boundary with
fractional winding modes  describes a configuration made of 
$2 \prod_b W^b $ lower dimensional $Dp$ branes.
This configuration can be obtained by  taking two
$Dp$ branes spaced along each
T-dualized direction $X^{t a}$ by  
$2\pi\left(\frac{1}{W^a}+ e_{(0)} a_{(0) a}- e_{(\pi)}
a_{(\pi)a}\right)$,
inscribing them into a $T^{d_\perp}$ torus with side lengths
$\frac{2\pi}{W^a}$  
and then tiling the $T^{d_\perp}$  torus with side lengths $2\pi$
using this configuration.
In particular the lowest energy open string configuration can be pictorially
portrayed as a sum of $\prod_b W^b$ configurations, each of which has a string
hanging between two $Dp$ in the same tile.
}
\COMMENTO{
BUH???
For example the part of the vertex describing 
an open tachyon emission in the case
of two perpendicular directions $a=1,2 $ of winding $W^1=2$ and $W^2=3$ 
is given by
\begin{eqnarray*}
V_T(z)
&\sim&
e^{i \frac{1}{2} X^1 +i \frac{2}{3} X^2 }
\left(\begin{array}{cc}
\sigma_+ & 0_2 \\
 0_2 & \sigma_+
\end{array}
\right)
\otimes
\left(\begin{array}{ccc}
0_2 & \sigma_+ & 0_2 \\
0_2 &  0_2 & \sigma_+ \\
\sigma_+ & 0_2 & 0_2
\end{array}
\right)
\end{eqnarray*}
}

In the general case the description is more complicated and involves a
twisting of the torus on which the open string lives. 
To find the open string description of the boundary, actually the
open string description of the boundary - boundary amplitude,
we can proceed as before and again
use the fact that the open string zero modes contribution to the one
loop free energy must be reproduced by the closed string zero modes
contribution to the boundary - boundary amplitude to fix the open
string periodicity. 
We assume that the open string ``homology''\footnote{
Actually $W$ contains homological information only in the directions parallel
  to the worldvolume.
} is given by
\begin{eqnarray}
X_{(open)}
&=&
X_{(open)}
+2\pi W s
\spazio
\forall s\in Z^d
\label{Dp_homology}
\end{eqnarray}
and we want to determine $W$.
If we parametrize the possible values of $n$ and $w$ entering the zero
modes part of the boundary (\ref{Dp0})
describing the $Dp$ using the quantities of the $D25$ boundary, 
\begin{eqnarray*}
w^m
=
W^{t\, m} l^{t\, m}
&\spazio&
w^a
= 
\alpha' (\gamma^{-T})^{a c} ( 2\pi F^t_{c m} W^{t\, m} l^{t\, m}
+2\pi F^t_{c b} W^{t\, b} l^{t\, b} )
\nonumber\\
n_m
=
2\pi F_{c m} w^{ m}
&\spazio&
n_a
=
2\pi F_{a b} w^{ b}
\end{eqnarray*}
where  $F$ is given by eq. (\ref{F_Dp}), $l^{t} \in \Z^d$ and
we write $2\pi F^t =W^{t\, -T}c^t W^{t\, -1}$ with $c^t\in Mat_d(\Z)$\footnote{
The $c^t_{i j}=-c^t_{j i}$ is the first Chern class along the 2-cycle parametrized
by $x^{t\,i}$ and $x^{t\,j}$.
} for the $D25$ field strength,
we can then compare the open string zero modes contribution to the one
loop free energy which is proportional to
$
\sum_{n\in\Z^d} e^{-\tau \alpha' (n+ W^T \Delta (e a) )^T W^{-1} 
\calG^{-1} W^{-T} (n+ W^T \Delta (e a) )}
$ 
to the closed string contribution using 
the  Poisson resummed open string contribution proportional to
$
\sum_{l\in\Z^d} e^{-  l^T W^{T} {\calG \over \alpha'} W l}
e^{i 2\pi l^T W^T \Delta(e a) }
$.

From this we can read the $Dp$ homology matrix which enters the definition
of $|Dp\rangle$ boundary state (\ref{Dp}) as
\begin{eqnarray}
W
&=&
\left(\begin{array}{cc}
P_\parallel W^{t} P_\parallel
&
0
\\
\alpha' P_\perp \gamma^{-T} W^{t\, -1} c^t P_\parallel
&
\alpha' P_\perp \gamma^{-T} W^{t\, -1} c^t P_\perp
\end{array}
\right)
\nonumber\\
&=&
\left(\begin{array}{cc}
 W^{t\,m} \delta^m_n
&
0
\\
\alpha' (\gamma^{-T})^{a c} {1\over W^{t\,c}} c^t_{c n} 
&
\alpha' (\gamma^{-T})^{a c} {1\over W^{t\,c}} c^t_{c b} 
\end{array}
\right)
\label{twisted_W}
\end{eqnarray}
so that the windings entering the boundary are 
$w=W l^t$ with  $l^t\in Z^d$
and the $y$ is connected to $y^t$ by
\begin{eqnarray}
y
&=&
\alpha' \gamma^{-T} y^t
\label{new_y}
\end{eqnarray}

We can then conclude that the open string sees a d-dimensional torus defined by
eq. (\ref{Dp_homology}) and eq. (\ref{twisted_W}) which is different
from the closed string torus defined  by eq. (\ref{X_closed_periodicity}).
Using eq. (\ref{Dp_homology}) we can read that the two $Dp$s wind 
$W^m=W^{t\, m}$
times in the direction $X^m$ parallel to the worldvolume exactly 
as the original $D25$ while they see twisted and shrunk perpendicular
directions.
The shrinking is $\frac{1}{W^{t\, a}}$ for any transverse direction
$X^a$
and therefore we get $\prod_b W^{t b}$ identical images on the closed
string torus. 
\COMMENTO{
These boundaries are the stringy way of  describing the same
configurations in pure (S)YM (see \cite{Taylor:1997dy} for a review).
}




\subsection{The degenerate case: $\Ft_{a b}=0$.}
As done in the previous section we begin with the simplest case, the
T-dual version of eq.s (\ref{simplest_B_1}) along $X^1$:
\begin{eqnarray}
\left( \hat n_1 -  \frac{p}{q}\hat n_2 \right) |B\rangle &=& 0
\nonumber\\
\left( \hat w^2 +  \frac{p}{q} \hat w^1  \right) |B\rangle &=& 0
\label{simplest_B_2_deg}
\end{eqnarray}
which can be derived from the boundary defining equations
$\calP_1 -\frac{p}{q} \calP_2 |B\rangle = \left(X^2 + \frac{p}{q}
X^1\right)' |B\rangle =0 $ which cannot be reduced to the standard form of
eq. (\ref{def_boundary}).
These equations suggest to consider the following canonical
transformation
\begin{eqnarray*}
\bar X = \Lambda X
&\spazio&
\bar \calP = \Lambda^{-T} \calP
\nonumber\\
\Lambda
=
\left(\begin{array}{cc}
r & s \\
p & q  \\
\end{array}
\right)
&,&
\Lambda^{-T}
=
\left(\begin{array}{cc}
q & -p \\
-s & r  \\
\end{array}
\right)
\in SL(2,\Z )
\end{eqnarray*}
We require $\Lambda\in  SL(2,\Z )$ since we do not want to change the
closed string theory and in particular its periodicity which is still
$\bar X \equiv \bar X + 2\pi s$ with $s\in Z^2$.
After this change of coordinate the boundary equations become trivially
\begin{eqnarray*}
\bar \calP_1 |B\rangle
& =&
\bar X^{' 2}|B\rangle
=0
\end{eqnarray*}
If we express the zero mode part of the boundary using the T-dual 
parameterization of the one given in eq. (\ref{D2}), i.e.
$\bar n_1=n_1, \bar w^1= w^1, \bar n_2=w^2, \bar w^2=n_2$
 we find that the
part of our interest can be written as
\begin{eqnarray*}
|B\rangle_{space\, zero\, modes}
&\sim& 
\sum_{l^t}
|\bar n_1 =0, \bar n_2 =-\frac{1}{W^{t\, 1}} l^{t\, 2},
\bar w^1 ={1\over W^{t\, 2}} l^{t\, 1}, \bar w^2 =0\rangle
\end{eqnarray*} 
from which we can deduce that
\begin{eqnarray*}
\bar X_{(open)} \equiv \bar X_{(open)} + 2\pi \bar W s
&\spazio&
\bar W
=
\left(\begin{array}{cc}
{1\over W^{t\, 2}} & 0 \\
0  & W^{t\, 1}  \\
\end{array}
\right)
\spazio
\forall s\in Z^2
\end{eqnarray*}
from which results clearly that the open string winds $\bar W^2=W^{t 1}\in\Z$
times in the direction $\bar X^2$ parallel to the world volume.

The general case is not treated since it is
more difficult to treat since it is harder to deal
with $SL(d,\Z)$ matrices which are needed to move the system in the
proper coordinates.

Finally we notice that we 
can find the same results using the following reasoning (more details will
be given elsewhere \cite{Pesando:2005mah}).
From the previous experience with
the emission vertexes for closed string states in open string
formalism \cite{Ademollo:1974fc,Frau:1997mq,Pesando:2003ww} we 
expect that the closed string tachyon emission vertex can
be written as 
$W_{T_c (o)} \sim :e^{i k_L^T X_L(z)}:\,
:e^{i k_R^T  X_R(\bar z)  )}:$ 
where  $X_L(z)$ and $X_R(\bar z)$ are the left and right moving
part of the {\sl open} string coordinate.
Since we know how $k_L$ and $k_R$  transform under a T-duality as they
are the closed string momenta and if we want a T-duality invariant
formalism,
we deduce immediately that the left and right moving part of the {\sl
  open} string coordinates must transform as the closed string ones,
i.e. as it follows from the extension of eq.s (\ref{Xt_X_relation}) to
an expression involving zero modes too\footnote{
These transformation rules were 
almost obtained in \cite{Sheikh-Jabbari:1999ac} in open
string formalism guessing the open string canonical transformations
(but missing the somewhat important $B$ dependence ) 
and in \cite{Dorn:1996an} using path integral formalism in the special
case of trivial homology and $F_{a b}=0$ where an isometry is present.
}:
\begin{eqnarray}
X_{L (open)} =  (P_\perp \gamma^{-T}E^{t\,T} +P_\parallel) X_{L (open)}^t
&\spazio&
X_{R (open)} =  - (P_\perp \gamma^{-T}E^{t} - P_\parallel) X_{R (open)}^t
\nonumber\\
 ~\label{Xt_X_open_relation}
\end{eqnarray}
which satisfy the boundary conditions we can read from the boundary
state defining eq.s (\ref{T-duality-def_boundary}).
In particular we get the same b.c. as in eq. (\ref{Neu tral_Xopen_b c})
\begin{eqnarray*}
&& G_{\mu \nu} X'^\nu
+ 2\pi\alpha'\, \dot X^\nu \calF_{\nu\mu} |_{\sigma=0,\pi}= 0
\end{eqnarray*}
in the non degenerate case where now $F$ and $E$ are related to $\Ft$ and
$E^t$ by eq.s (\ref{F_Dp}) and (\ref{E_Et})  respectively and 
\begin{eqnarray*}
\partial_\tau
\left[  X^{ a} - (\alpha' \gamma^{-T})^{a b} 2\pi \Ft_{b m} X^{    m}
\right]
|_{\sigma=0,\pi}
&=& 0
\nonumber\\
\left( G_{m i} -2\pi\alpha' \Ft_{m a} (\gamma^{-1})^{a b} G_{b i}
\right)X'^i 
+ \left( B_{m i} -2\pi\alpha' \Ft_{m a} (\gamma^{-1})^{a b} B_{b i}
\right)
\dot X^i 
&&
\nonumber\\
-  2\pi\alpha'\Ft_{m n} \dot X^n  
|_{\sigma=0,\pi}
&=&
 0
\end{eqnarray*}
in the case $\Ft_{a b}=0$. These equations reduce to (\ref{Neutral_Xopen_b
  c_Dp}) in the simpler case $\Ft_{a m}=0$.

\section{Conclusions.}
In this article we have shown it is possible to describe branes with
non trivial homology and non trivial but constant open background in
a non trivial but constant closed background on tori. 
An interesting and somewhat unexpected consequence is that the
physical Hilbert space of closed string theory is not enough and
therefore it must be extended with some ``twisted'' non physical
sectors in order to give a closed string representation of open string
states. This extension is dictated by the knowledge of the open string
spectrum and cannot be guessed, at least in an obvious way, just from
the closed string spectrum. 
This fact can have non trivial consequences in the
construction of branes in curved background using Ishibashi formalism.
Infact also the boundary state in flat space can be understood as a
weighted sum of Ishibashi states, therefore it can be necessary to
construct Ishibashi states using non physical (for the closed string) 
representations. 

Another point we discussed is the fact that open string T-duality does
not just amount to an exchange $\tau\leftrightarrow\sigma$ in presence
of a non trivial $B$ field but requires the open string fields to
transform as the closed string ones. Because of this $B$ and $F$
behave differently under T-duality. This fact can also simply
explained by noticing that $B$ transformation is ruled by closed
string  which does not know anything about the open field $F$.
As a further consequence it turns out that if we derive $Dp$ branes by
T-duality their geometrical embedding properties are only determined by
$F$.
In particular this is also true for type I string where only discrete
values of $B$ are allowed \cite{Bianchi:1997rf} since $B$ is not in the spectrum because T
duality rules are inherited from type IIB.

{\bf Acknowledgments}

\noindent
The author thanks M. Frau and  A. Lerda and 
in particular  M. Bill\'o and S. Sciuto for discussions.
This work supported in part by the European Community's 
RTN Program under contract Constituents, Fundamental Forces and
Symmetries of the Universe MRTN-CT-2004-005104.
It is also partially supported by the Italian MIUR under the program
``Teoria di Gauge,Gravit\'a e Stringhe''.

\newpage
\appendix
\section{Conventions.}
\label{conventions}
We define:
\begin{itemize}
\item 
WS metric signature: $\eta_{\alpha \beta }=(-,+)$; 
$\epsilon^{ 0  1}=-1$;
$\xi^0=\tau, \xi^1=\sigma$ 
\\ 
Space-time metric signature: $G_{\mu \nu}=(-,+,\dots, +)$;

\item
Indexes:
$D=26$,
$\mu,\nu,\dots=0,\dots D-1$,
$i,j,\dots=1,\dots,d=D-1$ are splitted into two sets of indexes: 
$ m,n,\dots$ for the directions along which we do not T-dualize,
$a,b,\dots$ for the directions which we T-dualize

\item 
$
z=e^{2(\tau_E+i\sigma)}, 
\bar z=e^{2(\tau_E-i\sigma)}~~,
$
with $\sigma\in[0,\pi]$

\item
All matrices have spacial only indexes, i.e.
$X= || X^i ||$, $p=|| p^i||$, $k=||k_i||$,
$\calG = ||\calG_{i j}||$,...

\item
$X^\mu$ , $F_{\mu \nu}$ are dimensionless 
while $G_{\mu \nu}$, $B_{\mu \nu}$ have the
  same dimension of $\alpha'$
\end{itemize}



\begin{thebibliography}{99}

\bibitem{DiVecchia:2005vm}
P.~Di Vecchia, A.~Liccardo, R.~Marotta and F.~Pezzella,
``On the gauge / gravity correspondence and the open/closed string duality,''
[arXiv:hep-th/0503156].


\bibitem{Murakami:2002yd}
K.~Murakami and T.~Nakatsu,
``Open Wilson lines as states of closed string,''
Prog.\ Theor.\ Phys.\  {\bf 110} (2003) 285
[arXiv:hep-th/0211232].

\bibitem{Pesando:2003ww}
I.~Pesando,
``Multibranes boundary states with open string interactions,''
[arXiv:hep-th/0310027].
\\
I.~Pesando,
``On the effective potential of the Dp Dp-bar system in type II  theories,''
Mod.\ Phys.\ Lett.\ A {\bf 14} (1999) 1545
[arXiv:hep-th/9902181].

\bibitem{Pesando:2005mah}
I.~Pesando,
``Open and closed string vertex operators for a neutral string, ''
in preparation


\bibitem{Uranga:2003pz}
A.~M.~Uranga,
``Chiral four-dimensional string compactifications with intersecting
D-branes,''
Class.\ Quant.\ Grav.\  {\bf 20} (2003) S373
[arXiv:hep-th/0301032].

\bibitem{Kiritsis:2003mc}
E.~Kiritsis,
``D-branes in standard model building, gravity and cosmology,''
Fortsch.\ Phys.\  {\bf 52} (2004) 200
[arXiv:hep-th/0310001].


\bibitem{Arfaei:1999jt}
H.~Arfaei and D.~Kamani,
``Branes with back-ground fields in boundary state formalism,''
Phys.\ Lett.\ B {\bf 452} (1999) 54
[arXiv:hep-th/9909167].

\bibitem{Kamani:2002bb}
D.~Kamani,
``Closed superstring in noncommutative compact spacetime,''
Mod.\ Phys.\ Lett.\ A {\bf 17} (2002) 2443
[arXiv:hep-th/0212088].

\bibitem{Kamani:2002ib}
D.~Kamani,
``Space-time symmetries, T-duality and gauge theory,''
Phys.\ Lett.\ B {\bf 541} (2002) 406
[arXiv:hep-th/0205189].

\bibitem{Kamani:2001cg}
D.~Kamani,
``General target space duality and its effects on D-branes,''
Nucl.\ Phys.\ B {\bf 601} (2001) 149
[arXiv:hep-th/0104089].


\bibitem{Giveon:1988tt}
A.~Giveon, E.~Rabinovici and G.~Veneziano,
``Duality In String Background Space,''
Nucl.\ Phys.\ B {\bf 322} (1989) 167.

\bibitem{Bianchi:2005yz}
M.~Bianchi and E.~Trevigne,
``The open story of the magnetic fluxes,''
arXiv:hep-th/0502147.

\bibitem{Abouelsaood:1986gd}
A.~Abouelsaood, C.~G.~.~Callan, C.~R.~Nappi and S.~A.~Yost,
``Open Strings In Background Gauge Fields,''
Nucl.\ Phys.\ B {\bf 280} (1987) 599.
%

\bibitem{Chu:1999gi}
C.~S.~Chu and P.~M.~Ho,
``Constrained quantization of open string in background B field and
noncommutative D-brane,''
Nucl.\ Phys.\ B {\bf 568} (2000) 447
[arXiv:hep-th/9906192].

\bibitem{Seiberg:1999vs}
N.~Seiberg and E.~Witten,
``String theory and noncommutative geometry,''
JHEP {\bf 9909} (1999) 032
[arXiv:hep-th/9908142].

\bibitem{DiVecchia:1997pr}
P.~Di Vecchia, M.~Frau, I.~Pesando, S.~Sciuto, A.~Lerda and R.~Russo,
Nucl.\ Phys.\ B {\bf 507} (1997) 259
[arXiv:hep-th/9707068].

\bibitem{DiVecchia:1999rh}
  P.~Di Vecchia and A.~Liccardo,
  NATO Adv.\ Study Inst.\ Ser.\ C.\ Math.\ Phys.\ Sci.\  {\bf 556} (2000) 1
  [arXiv:hep-th/9912161].



\bibitem{Ademollo:1974fc}
M.~Ademollo, A.~D'Adda, R.~D'Auria, E.~Napolitano, P.~Di Vecchia, F.~Gliozzi and S.~Sciuto,
``Unified Dual Model For Interacting Open And Closed Strings,''
Nucl.\ Phys.\ B {\bf 77} (1974) 189.

\bibitem{Frau:1997mq}
M.~Frau, I.~Pesando, S.~Sciuto, A.~Lerda and R.~Russo,
``Scattering of closed strings from many D-branes,''
Phys.\ Lett.\ B {\bf 400} (1997) 52
[arXiv:hep-th/9702037].

\bibitem{Sheikh-Jabbari:1999ac}
M.~M.~Sheikh-Jabbari,
``A note on T-duality, open strings in B-field background and canonical
transformations,''
Phys.\ Lett.\ B {\bf 474} (2000) 292
[arXiv:hep-th/9911203].

\bibitem{Dorn:1996an}
H.~Dorn and H.~J.~Otto,
``On T-duality for open strings in general abelian and nonabelian gauge field
backgrounds,''
Phys.\ Lett.\ B {\bf 381} (1996) 81
[arXiv:hep-th/9603186].

\bibitem{Bianchi:1997rf}
M.~Bianchi,
``A note on toroidal compactifications of the type I superstring and  other
superstring vacuum configurations with 16 supercharges,''
Nucl.\ Phys.\ B {\bf 528} (1998) 73
[arXiv:hep-th/9711201].


\end{thebibliography}
\end{document}